# Comparative modeling studies of TSDC: investigation of Alpha- relaxation in Amorphous polymers

A. E. Kotp[1,2]

[1]*Physics Dept., Faculty of Science, Mansoura University, Mansoura, Egypt.*
[2]*Physics Dept., Faculty of Science and Humanitarian Studies, Alkahrj University, Alkahrj, KSA*

**Abstract:**

A model to investigate Thermally Stimulated Depolarization Current (TSDC) peak parameters using the dipole-dipole interaction concept is proposed by the author in this work. The proposed model describe the (TSDC) peak successfully since it gives a significant peak parameters (i.e. Activation energy (E) and the per- exponential factor ($\tau_o$) in addition to the dipole-dipole interaction strength parameter ($d_i$). Application of this model to determine the peak parameters of polyvinyl chloride(PVC) polymer is presented . The results show how the model fit the experimental thermal sampling data. Finally the results are compared to the well know techniques; the initial rise method (IR), the half width method (HW) in addition to the Cowell and Woods analysis..

Introduction:

Investigation of molecular motion in polymers grasping more attention, however, with the aid of modeling and experimentation we take an insight through the molecular motion. The most well known study of molecular motion is the study relaxation processes take place inside the polymers. In these studies almost two techniques are used the first is dielectric spectroscopy (DS) and the other is the thermally stimulated depolarization current (TSDC).

The polymer as a modeling system is very large and complicated molecular system which results in using the phenomenology of the relaxation process ( i. e , without considering the objective reality of the process).

In this work a model was presented, which depend on the polarization phenomena, which take place in the relaxation process during the TSDC

experiments, in addition to the dipole-dipole interaction concept which take place between the oriented dipoles during the orientation process. The model provide a data to investigate the contribution of the dipole-dipole interaction to the whole TSDC curve.

Theoretical Model Development:

The mathematical model was made by assuming the contribution of the dipole-dipole interaction is affecting only the width of the TSDC peak, which leads to assuming further that the error due to the dipole-dipole interaction in the TSDC curve is distributed normally on the curve. By considering the Gaussian distribution according to the local limit theory[1] it is found that the error due to the dipole-dipole interaction is given by a correction factor:

$$C_f = \frac{1}{d_i \sqrt{\pi}} \int_{-\infty}^{\infty} \exp(\frac{-E - E_o}{kT}) \exp(\frac{(-E - E_o)^2}{d_i}) \exp(\frac{-1}{\beta} \int_{T_1}^{T_2} (\frac{1}{\tau(T)})(\frac{1}{\tau_{Eo}(T)}) dT) \quad (1)$$

Where, ($E_o$) is the center of the energy distribution, ($d_i$) is its width, (ß=dT/dt) is heating rate, (E) is the activation energy and $\tau(T)$ is the relaxation function.

Multiplying the correction factor ($C_f$) with the TSDC curve model given by:

$$I(T) = \frac{(AP_o)}{\tau} \exp((\frac{-1}{\beta}) \int_{T1}^{T2} (\frac{1}{\tau} dT)) \quad (2)$$

Where, (A) is sample area, ($P_o$) is the initial polarisation and $\tau(T)$ is the relaxation model used, gives the corrected TSDC curve given by:

$$I_{corr}(T) = C_f * I(T) \quad (3)$$

The theoretical correction factor ($C_f$) is studied very intensively in order to get a complete picture about the correction factor behavior along with all variables and constants using MathCAD.

The Dipole-dipole interaction (DDIA) model was developed using an approach depending on the Arrhenius relaxation as a relaxation function.

Firstly inserting the Arrhenius relaxation model $\tau(T)$ given by the well known equation:

$$\tau(T) = \tau_o \exp(\frac{E}{kT}) \qquad (4)$$

in the correction factor equation we have:

$$C_f = \frac{1}{(d_i \sqrt{\pi})} \int_{-\infty}^{\infty} \exp(\frac{(-E-E_o)}{kT}) \exp(\frac{(-E-E_o)^2}{d_i}) \exp[\frac{-1}{\beta} \int_{T1}^{T2} \exp(\frac{-E}{(kT)}) \exp(\frac{-E_o}{(kT)}) dT] dE \qquad (5)$$

Also we can do the same for equation (2) to become:

$$I(T) = \frac{(AP_o)}{\tau_o} \exp(\frac{-E}{kT}) \exp(\frac{-1}{(\beta \tau_o)} \int_{T1}^{T2} \exp(\frac{-E}{kT}) dT) \qquad (6)$$

Using equations (3),(5), and (6) we have a model to calculate the TSDC peak parameters based on the dipole-dipole interaction concept in Arrhenius relaxation mode which we shall call it (DDIA) model.

After the analysis of the correction factor ($C_f$) a Fortran 95 program was made in order to compute equation **(3).**

The improper integral over the energy (E) in equation (5) was approximated[2] the integral over T was calculated using the Simpson Composite method[3] where it is found to be more accurate for the exponential function. The TSDC peak parameters was calculated by Fitting this model using least square method to the experimental data .

Results and discussion:

In order to investigate the the dipole-dipole interaction model in the Arrhenius mode  DDIA model applicability an experimental data peaks obtained from Poly(vinyl chloride)PVC, using the thermal sampling technique was used. Furthermore to compare the model calculated peak parameters (i.e. the energy (E) and pre-exponential factor ($\tau_o$)) two known peak shape methods was used;  the initial rise method (IR)[4,5], the half width method (HW)[6] in addition to the Cowell and Woods[7] analysis.

In order to see the quality of fitting of the DDIA model with the experimental data the calculated TSDC points was plotted against the experimental data. Furthermore, the DDIA model fitting was compared to the Cowell-Wood method.

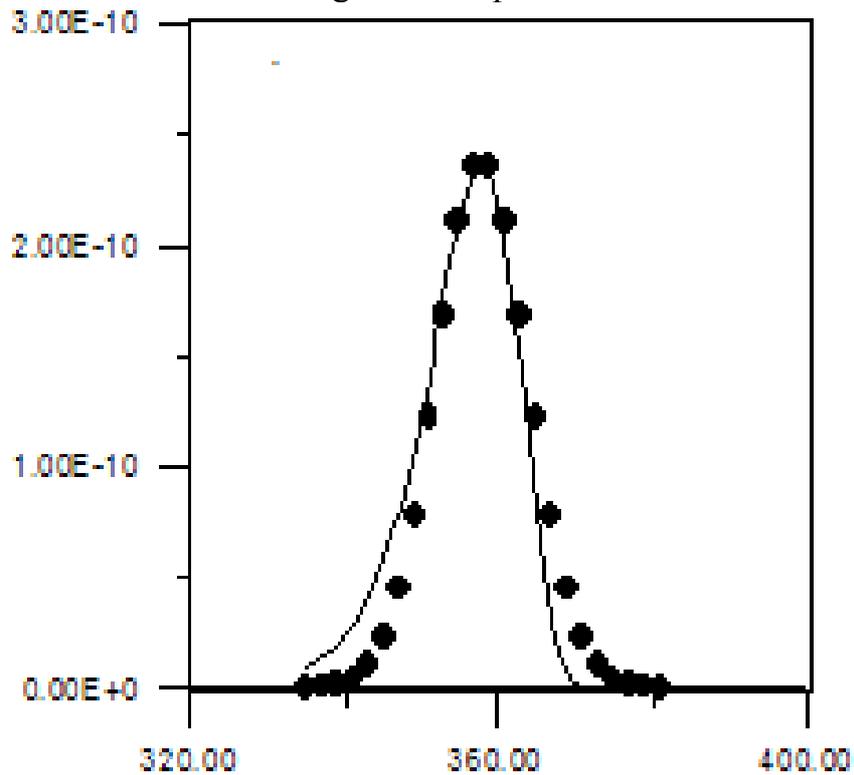

Figure 1: The fitting of Cowell-Wood method and the experimental Thermal Sampling data for peak no. 4 in table 1.

Figure 1. shows the fitting of the experimental data and the calculated data using Cowell-Wood method. It shows the poor fitting specially in the high and low temperature sides. Figure.2 shows the fitting of experimental data and the calculated data using DDIA model. It shows that the fitting is much more correlated to the experimental data which indicate that the dipole-dipole interaction is affecting the TSDC current peak, which was neglected before in the analysis of the TSDC current peak.

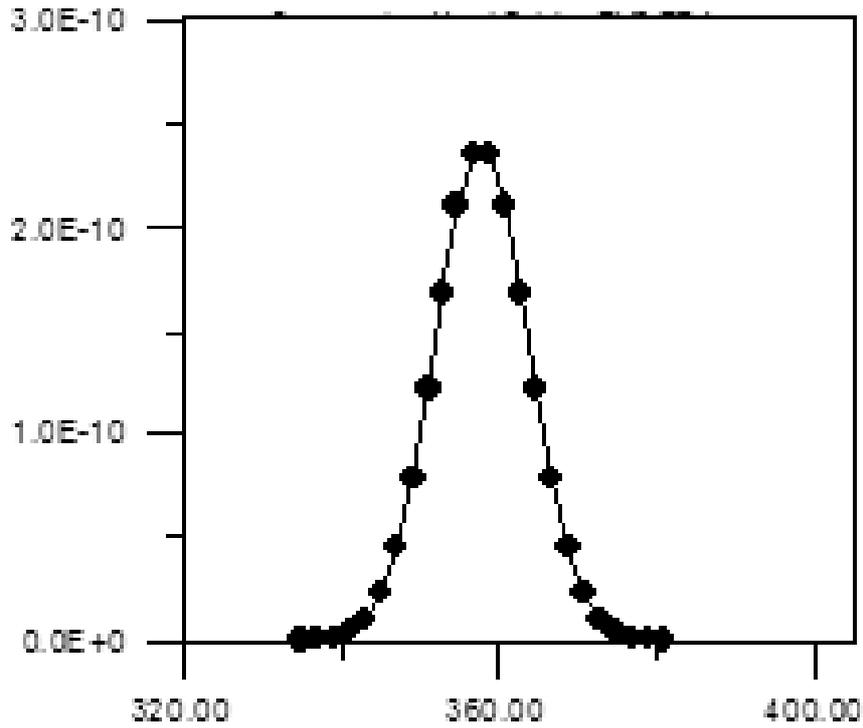

Figure 2: The fitting of DDIA model and the experimental Thermal Sampling data for the TS peak no. 4 in table 1.

Table 1. shows the dipole-dipole interaction model in Arrhinus mode (DDIA) calculated peak parameters (E), ($\tau_o$) in addition to the dipole-dipole interaction strength parameter ($d_i$). Both the activation energy (E) and dipole-dipole interaction strength ($d_i$) is introduced as (Kj/mole) and the pre-exponential factor ($\tau_o$) is introduced in seconds for PVC polymer material.

Table(1): The calculated thermal sampling peak parameters using the dipole-dipole interaction model in Arrhenius mode (DDIA) in addition to the initial rise (IR) method, half width (HW) method, and Cowell-Wood (CW) method.

| Method | | | IR | | HW | | CW | | DDIMA | | |
|---|---|---|---|---|---|---|---|---|---|---|---|
| Mat. | Ts. No | $T_m$ (K) | E (Kj/mole) | $\tau_o$(sec.) | E (Kj/mole) | $\tau_o$(sec.) | E (Kj/mole) | $\tau_o$(sec.) | E (Kj/mole) | $\tau_o$(sec.) | $d_i$ (Kj/mole) |
| PVC | 1 | 345 | 117 | $3.131 \times 10^{-16}$ | 191 | $9.702 \times 10^{-28}$ | 192 | $6.503 \times 10^{-28}$ | 125 | $1.410 \times 10^{-17}$ | 8.58 |
| | 2 | 347 | 129 | $5.734 \times 10^{-18}$ | 194 | $6.503 \times 10^{-28}$ | 196 | $2.392 \times 10^{-28}$ | 127 | $1.044 \times 10^{-17}$ | 8.68 |
| | 3 | 351 | 153 | $2.869 \times 10^{-21}$ | 290 | $9.102 \times 10^{-42}$ | 194 | $1.072 \times 10^{-27}$ | 153 | $3.171 \times 10^{-21}$ | 8.39 |
| | 4 | 357 | 160 | $4.292 \times 10^{-22}$ | 296 | $2.742 \times 10^{-42}$ | 281 | $2.448 \times 10^{-37}$ | 157 | $1.055 \times 10^{-21}$ | 7.04 |
| | 5 | 361 | 192 | $1.182 \times 10^{-26}$ | 346 | $3.783 \times 10^{-44}$ | 266 | $2.448 \times 10^{-37}$ | 201 | $1.075 \times 10^{-28}$ | 8.21 |
| | 6 | 363 | 298 | $1.005 \times 10^{-41}$ | 306 | $5.535 \times 10^{-43}$ | 242 | $1.330 \times 10^{-33}$ | 203 | $5.324 \times 10^{-28}$ | 8.29 |
| | 7 | 365 | 208 | $1.313 \times 10^{-28}$ | 310 | $3.040 \times 10^{-43}$ | 300 | $9.102 \times 10^{-42}$ | 198 | $3.934 \times 10^{-27}$ | 8.11 |
| | 8 | 379 | 109 | $1.705 \times 10^{-13}$ | 176 | $5.809 \times 10^{-23}$ | 154 | $8.599 \times 10^{-20}$ | 111 | $8.467 \times 10^{-14}$ | 7.52 |

The model calculated results shows that the DDIA data give a significant figures for the pre-exponential factor ($\tau_o$) while the initial rise method, the half width and Cowell-Wood methods give non significant figures. This give an indication that the DDIA model gives more correction to the calculated relaxation time. Furthermore, the calculated activation energy ($E_a$) using the DDIA model is less in magnitude than those calculated using the other methods which gives more significant values for the activation energy. Figure3 shows a comparison of the calculated activation energy using the four different methods, IR, HW, CW, and DDIA respectively. In this figure it is found that the activation energies calculated using the IR method and DDIA are closely correlated except in TS peak no. 6, the

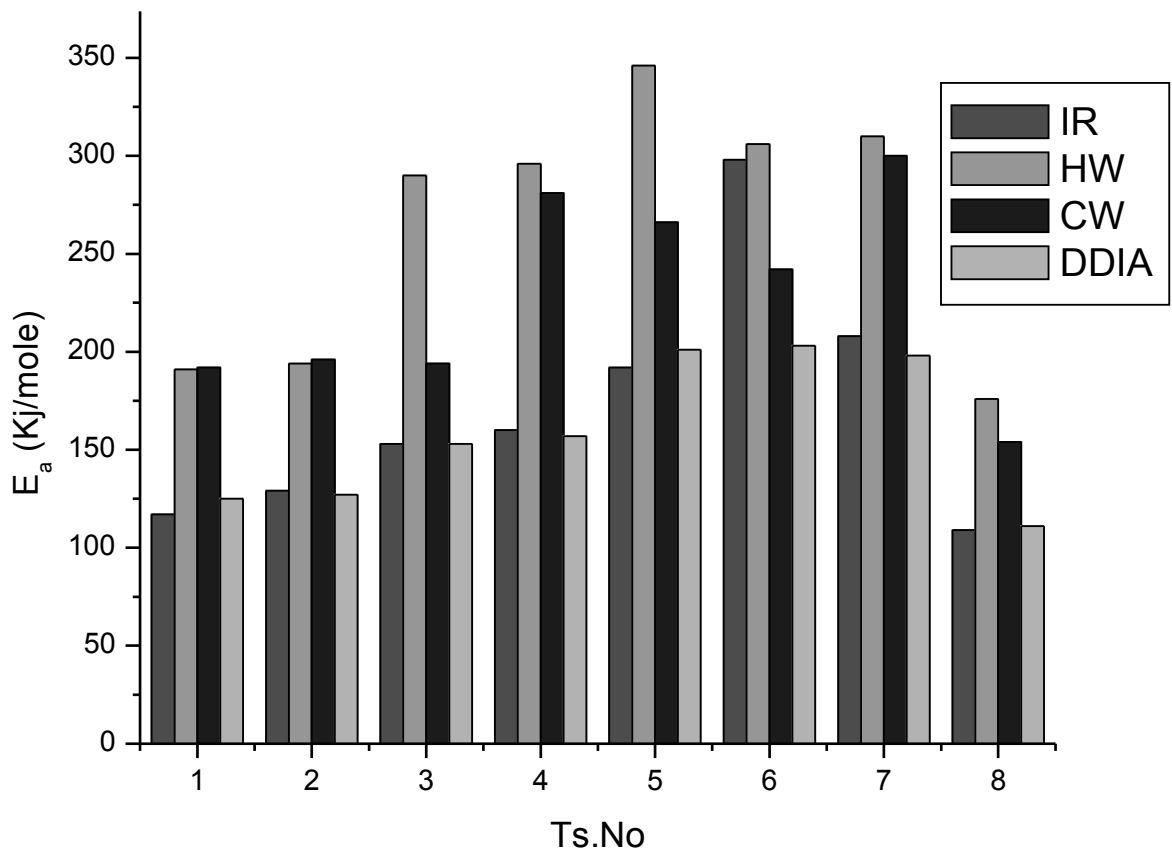

Figure3: The activation energy ($E_a$) values obtained using the three different methods IR, HW, CW, and DDIA model.

calculated activation energy using the IR method gives a very high value 298Kj/mole ,where as, DDIA model gives a more significant value 203Kj/mole. These results indicate that the DDIA model correct the errors in the peak shape methods (IR, HW) and also gives more correct values than the CW method. On the other hand it is found that the CW method is much correlated to the HW method except in TS peak 3 and 5 they give much far values.

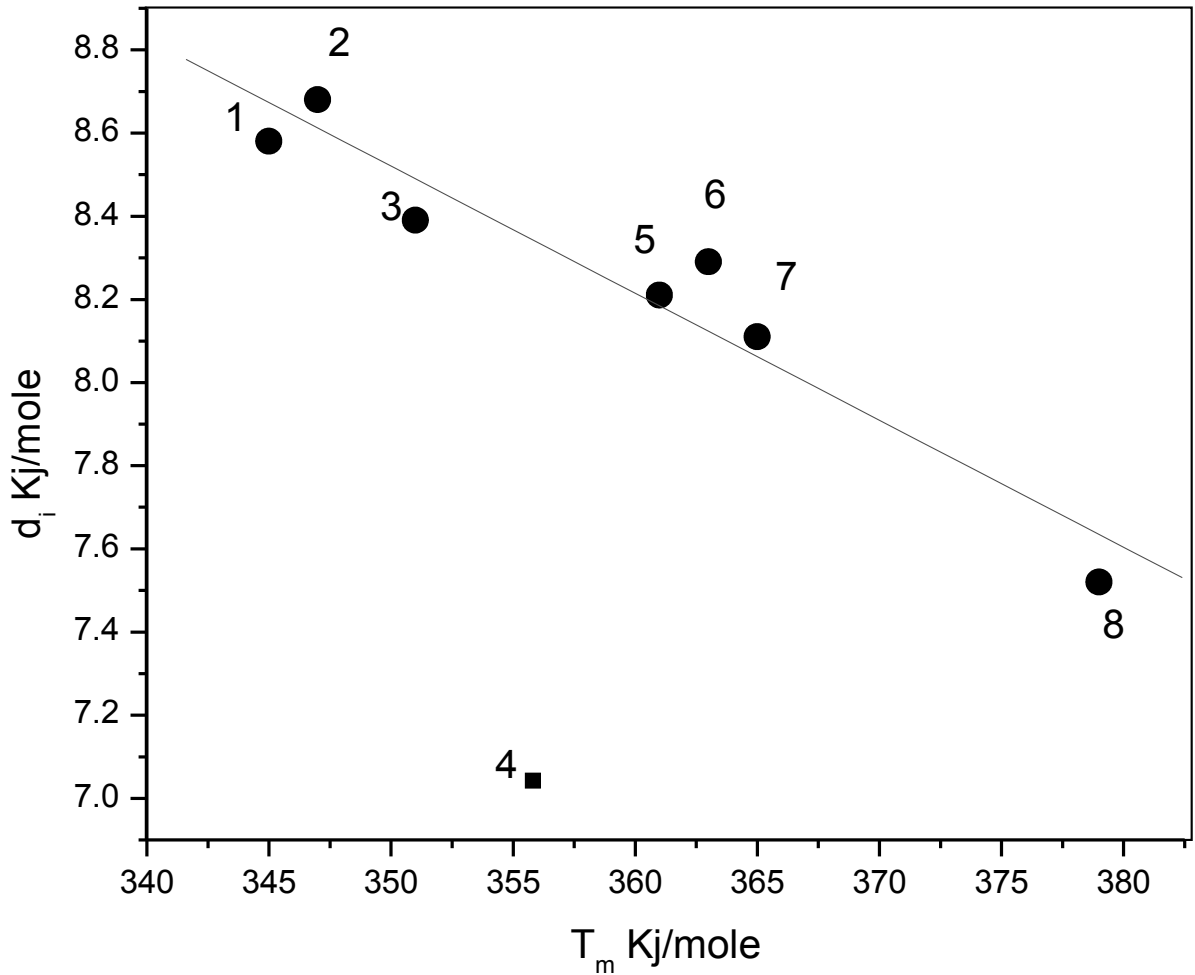

Figure4: The relationship between the calculated dipole-dipole interaction parameter ($d_i$) and the maximum temperature ($T_m$).

Figure4 shows the dipole-dipole interaction parameter ($d_i$) against the maximum temperature ($T_m$) which show that this parameter depend linearly, except TS peak no.4, on the temperature.

Conclusion:

The dipole-dipole interaction model correct the high and low temperature sides disagreement between experimental data and the calculated data using the CW method. Which mean that the dipole-dipole interaction affect the TSDC peak by increasing its width.

The dipole-dipole interaction model in Arrhenius mode DDIA give more significant peak parameters values than the other methods used especially in the calculation of the pre-exponential factor ($\tau_o$).

Using the dipole-dipole interaction model in Arrhenius mode DDIA lead to computation of the dipole-dipole interaction parameter ($d_i$) which give an indication for the magnitude of this parameter in polymers.